\documentclass[prd,twocolumn,nofootinbib,amsmath,amssymb]{revtex4-2}
\usepackage{graphicx}
\usepackage{dcolumn}% Align table columns on decimal point
\usepackage{bm}% bold math
\usepackage{latexsym}
\usepackage{amssymb}
\usepackage{amsmath}
\usepackage{subfigure}
\usepackage{longtable}
\usepackage{multirow}
\usepackage[running]{lineno}
\usepackage{hyperref}
\begin{document}

	\title{Froissaron and Maximal Odderon with spin-flip\\ 
		in $pp$ and  $\bar pp$ high energy elastic scattering}% Force line breaks with \\

\author{N. Bence,} 
\affiliation{%	
	Uzhgorod National University, 88017 Universitetska street, 14, Uzhgorod, \\
e-mail: bencenorbert007@gmail.com}
\author{A. Lengyel, Z. Tarics,}
\affiliation{%	
	{Institute of Electron Physics of NAS of Ukraine,\\ 88017 Universytetska street. 21, Uzhgorod\\
e-mails: alexanderlengyel39@gmail.com;  tarics1@rambler.ru}
}
\author{E. Martynov,  G. Tersimonov, }  
\affiliation{%
	Bogolyubov Institute for Theoretical Physics of NAS of Ukraine,\\ 03143, Metrologicheskaja st. 14b, Kiev-143, Ukraine.\\
	e-mails: martynov@bitp.kiev.ua;  tersimonov@bitp.kiev.ua}

%\date{\today}

\begin{abstract}
	We assume that the scattering amplitude is represented by Froissaron, Maximal Odderon as well as by standard Regge poles. From the fit to the data of $pp$ and  $\bar pp$ scattering at high energy and not too large momentum transfers we found that this model taking into account the spin is available to describe not only the  differential, total cross section and $\rho$, but also the  existing  experimental data on polarization.
\end{abstract}

\maketitle

\section*{Introduction}\label{sec:Introd}

The polarization data available at present for proton-proton scattering at relatively high energies are compatible with the $s$-channel helicity conservation \cite{Leader, Bultimore, ELNS} and show that the polarization contribution decreases with increasing energy as it is expected in terms of Regge pole exchanges. The data are insufficiently precise to provide unambiguous conclusion at the higher energy, although a contribution from the helicity-flip amplitude cannot be excluded \cite{Leader, TT, DDLN, Predazzi}. The existing polarized experiments \cite{Fidecaro-1, Fidecaro-2, Kline-1, RHIC} at energy lower than LHC  allow one to study spin properties of proton-pomeron  vertex at intermediate energies while conclusions about  spin properties of amplitudes at high $t$ are derived as rule  from model extrapolations.   

The main conclusion was made that  pomeron exchange is expected to produce the observed small spin effects. 

Despite  a long time interest to spin effects physics in hadron interaction the available data set is not rich in a soft kinematic region. Firstly, the main part of experiments was performed for meson-nucleon and electron-proton interactions. Secondly, these measurements were made at low and intermediate energies.  Nevertheless, the available data stimulate permanently the theoretical and phenomenological studies of spin phenomena in the regions of soft and hard kinematic, where different approaches (nonperturbative  approches from early \cite{BLSNM-T} to relatively later and recent \cite{Predazzi, CPS, Selyugin-1, Selyugin-2, BZK-MK}, and perturbative QCD)  are exploring \cite{ GalKur}.  

Unfortunately, the experimental data on spin observable quantities at the highest energies are almost absent, though they may provide additional information about  helicity amplitudes properties . As we known only the data at FNAL energies \cite{Fidecaro-1,Fidecaro-2,Kline-1} and the relatively recent data from RHIC \cite{RHIC} (however, at very small $t$, in Coulomb-nuclear interference region) are available.   

However,  now thanks to a series of CERN TOTEM proton-proton experiments \cite{TOTEM-7,TOTEM-8,TOTEM-13,TOTEM-2-76} we possess the long-awaited complete set of the precise data for testing various models over a wide  energy and momentum transfer range. Proton-proton (antiproton) scattering is the unique possibility to investigate crossing-odd contribution and its spin-flip properties. 
 
The Froissaron and Maximal Odderon model (the FMO model) \cite{LN, MN-1,MN-2,MN-3} has recently successfully described such an extended set of experimental data. Therefore, the idea to extend the application of this model to consider some of the spin effects, for instance,  polarization data of $pp$ scattering at relatively high energy naturally arises. This task has been around for a long time, although the  attempts to solve it are rarely undertaken, as we noted above, due to a lack of sufficiently precise data at the highest energies. 

The main interest of the present paper is concentrated, after new measurements at the LHC, on the questions ''Do Froissaron and Maximal Odderon flip spin?\,`` and  ''How big are crossing-even and crossing-odd  spin-flip amplitudes as compared to  the corresponding spin-non-flip-amplitudes? \!\!``.  We examine the issue in the framework of the  Froissaron and Maximal Odderon model.  In Froissaron and Maximal Odderon approach we restrict ourselves to taking into account the contribution of two amplitudes: the spin-non-flip and the single spin-flip ones.

The paper is organized as follows. In the next section the main formulas and general discussion of the problems are presented.  In Section \ref{sect:modelA} we describe the Simplified FMO amplitude. 
 In the  Section \ref{sect:modelB} we discuss the original FMO model  with spin-flip amplitudes. The results of comparison of both models  with the data at $\sqrt{s}>$19 GeV and at intermediate $t$ is given  in the Section \ref{sect:fresults}.

To avoid the inevitable increase in the number of parameters, we use similarly to many authors \cite{Leader, Bultimore, ELNS, TT, DDLN, Predazzi, CPS, Selyugin-1, Selyugin-2} some simplified assumptions  about  the spin-non-flip and spin-flip amplitudes and their relationship.

\section{Definitions and the FMO approach}

\subsection{Helicity amplitudes, observables}

Generally the proton-proton and antiproton-proton scattering amplitude reads as 
\begin{equation}\label{eq:def-A-0}
A^{ pp}_{\bar pp}(s,t)=A^{(+)}(s,t) \pm A^{(-)}(s,t).
\end{equation}	
In this model we used the following normalization of the spin-averaged physical amplitudes 
\begin{equation}\label{eq: observ}
\begin{array}{ll}
\sigma_t(s)&=\dfrac{1}{\sqrt{s (s-4m^2 )}}\text{Im} A(s,0), \\
% \rho=\dfrac{\Re e F(s,0)}{\Im m F(s,0)}, \quad 
\dfrac{d\sigma_{el}}{dt}&=\dfrac{1}{64\pi ks(s-4m^2)}|A(s,t)|^2
\end{array}
\end{equation}
where $k=0.3893797\,\, \text{mb}\cdot\text{GeV}^2$. With this normalization the amplitudes have dimension $\text{mb}\cdot\text{GeV}^2$ and all the couplings are given in millibarns.

Taking into account the spin degrees of freedom in nucleon-nucleon elastic scattering still as a not well defined and quite complicated procedure. There are no  strict and consequent methods to construct 5 independent helicity amplitudes for elastic nucleon-nucleon interaction:  
\begin{equation}\label{eq:H-ampl}
\begin{aligned}
\Phi_1(s,t) = &\langle++|\mathcal{T} |++\rangle,\\
\Phi_2(s,t)= &\langle++|\mathcal{T} |--\rangle,\\
\Phi_3(s,t)= &\langle+-|\mathcal{T} |+-\rangle,\\
\Phi_4(s,t)= &\langle+-|\mathcal{T} |-+\rangle,\\
\Phi_5(s,t)= &\langle++|\mathcal{T} |+-\rangle.\\
\end{aligned}
\end{equation} 

Only various phenomenological models for these amplitudes have been constructed and analyzed. 
Each of these amplitudes has crossing-even and crossing-odd components $\Phi_i(s,t), \quad i=1,2,...,5$
\begin{equation}\label{eq:def-Phi-0}
\Phi^{ pp}_{\bar pp}(s,t)=\Phi^{(+)}(s,t) \pm \Phi^{(-)}(s,t).
\end{equation}	

For the spin-averaged initial protons (antiprotons) the following equations for the total and differential cross sections and polarization are used
\begin{equation}\label{eq:norm-stot}
\sigma_{tot}=\dfrac{1}{\sqrt{s(s-4m^2)}}\text{Im}\left ( \Phi_1+\Phi_3 \right ),
\end{equation}
\begin{equation}\label{eq:norm-dsdt}
\begin{aligned}	
\dfrac{d\sigma_{el}}{dt}&=\dfrac{1} {16\pi k s(s-4m^2)}\\
&\times \left ( |\Phi_1|^2+|\Phi_2|^2+|\Phi_3|^2+|\Phi_4|^2 +4|\Phi_5|^2 \right ),
\end{aligned}
\end{equation}
\begin{equation}\label{eq:Pol}
P(t)=-2\dfrac{\text {Im}[ (\Phi_1+\Phi_2+\Phi_3-\Phi_4) \Phi^*_5 ]}{|\Phi_1|^2+|\Phi_2|^2+|\Phi_3|^2+|\Phi_4|^2 +4|\Phi_5|^2}.
\end{equation}

Therefore in what follows we consider the case with $\Phi_2=\Phi_3=\Phi_4=0$, assuming in accordance with a common opinion that other spin amplitudes are small at  high energies \cite{Bultimore, Leader, DDLN, Predazzi, CPS, Selyugin-1, Selyugin-2}.  

\subsection{FMO approach}

Let's comment shortly the Foissaron and Maximal Odderon approach to elastic proton-proton (antiproton) elastic scattering. Two realizations  of it are explored in the paper. 

Historically it was formulated in  \cite{LN} and applied \cite{MN-1,MN-2,MN-3} in description of  high energy data on $pp$ and $\bar pp$ scattering including  the newest TOTEM data at 13 TeV. 

The  maximal growth of the even-under-crossing amplitude $A^{(+)}$ allowed by unitarity: 
\begin{equation}\label{eq:Froissaron}
	A^{(+)}(s,t=0)\propto is\ln^2\tilde s, \qquad \tilde s=-is/s_0, \quad s_0=1\,\,\text{GeV}^2  
\end{equation}
lead to
\begin{equation}\label{eq:stot}
	\sigma\propto \ln^2(s/s_0).
\end{equation}
The corresponding maximal behavior of the odd-under-crossing amplitude $A^{(-)}$
\begin{equation}\label{eq:Odderon}
	A^{(-)}(s,t=0)\propto s\ln^2\tilde  s
\end{equation}
leads via the optical theorem  to the difference of the antiparticle-particle and particle-particle total cross sections
\begin{equation}\label{eq:delstot}
	\Delta \sigma \propto \ln(s/s_0)
\end{equation}
which grows, in absolute value, with energy. However, the sign of $\Delta\sigma$ is not fixed by general principles.

Such a behaviour \eqref{eq:stot} is often referred to as ``Froissaron'', while the behaviour (\ref{eq:Odderon}) is termed as ``Maximal Odderon''. At $t=0$, they  corresponds in the $j$-plane to a triple pole located at $j=1$. It was shown in \cite{MN-1} that Froissaron and Maximal Odderon model describes perfectly all the forward scattering TOTEM data \cite{TOTEM-8, TOTEM-13, TOTEM-2-76}, including  the surprisingly small value of the ratio $\rho$ at $\sqrt{s}=13$ TeV. Moreover, fit to the data at $t = 0$ of the model with a more general form of FMO \cite{MN-2}  shows that one returns to the solution with Froissaron and Maximal Odderon. Extension of the FMO model for $t\neq 0$ was suggested in \cite{MN-3} and led to quite good fit in the region 
$\sqrt{s}\geq 5$ GeV at $t=0$ and $19\, \text{GeV} \sqrt{s} \leq 13\,\text{TeV}$ at $0< |t|\leq 5\, \text{GeV}^2$. Thus, it would be interesting to consider at least some of the spin effects within this  approach. 

\section{The models for  spin-non-flip and spin-flip amplitudes}\label{sect: simple model}

As noted above, our aim is to study the spin-flip effects, assuming that  they are determined mainly by the $\Phi_5$ term. 
Therefore, we apply here the assumption here the  assumption  $\Phi_2=\Phi_3=\Phi_4=0$ to the models considered.  If $\Phi_3\neq 0$ but has the same functional form as $\Phi_1$, then just redefinition of the couplings in  $\Phi_1$ can be applied.

\subsection{ Simplified FMO model (Model A)}\label{sect:modelA}

Only the main terms of Froissaron and Maximal Odderon amplitudes in FMO model \cite{MN-3} are taken into account in the model A.  At $t=0$ they correspond to triple poles in the $j$-plane. This truncated crossing-even and crossing-odd terms of FMO model are noted as $P^{(T)}_{i}(s,t)$ and $O^{(T)}_{i}(s,t)$, respectively.

Thus, in the model A the crossing-even and crossing-odd  parts of the both spin amplitudes important at high energy  are chosen as follows
\begin{equation}\label{eq;def-A-1}
	\begin{aligned}
\Phi^{(\pm)}_{i}(s,t)&=P_i(s,t)+R^f_i(s,t)\pm[O_i(s,t)+R_i^{\omega}(s,t)],\\
%\end{equation}
%\begin{equation}\label{eq:P}
P_{i}(s,t)&=P^{(T)}_{i}(s,t)+P^{(D)}_{i}(s,t)+P^{(S)}_{i}(s,t),\\
%\end{equation}
%\begin{equation}\label{eq:O}
O_{i}(s,t)&=O^{(T)}_{i}(s,t)+O^{(D)}_{i}(s,t)+O^{(S)}_{i}(s,t),
\end{aligned}
\end{equation}
where $i=1,5$. 
The full explicit expressions for the model A are given in Appendix  \ref{sect:sfmo}. 

Spin-flip amplitudes are written in the model A in the following form
\begin{equation}\label{eq:AP-5}
	\begin{aligned}
P_5^{(T,D,S,f)}(s,t&)=\dfrac{\sqrt{-t}}{2m}\lambda_{+}(t)P_1^{(T,D,S,f)}(s,t),\\  
O_5^{(T,D,S,\omega)}(s,t)&=\dfrac{\sqrt{-t}}{2m}\lambda_{-}(t)O_1^{(T,D,S,\omega)}(s,t), \\ P_i^f&\equiv R_i^f,\quad O_i^\omega\equiv R_i^\omega.
\end{aligned}
\end{equation}
For $\lambda_\pm(t)$ we try three variants in the fit to experimental data: 
\begin{equation} \label{eq:lambda}
\begin{aligned}	
\text{V.1} \quad   \lambda_\pm(t)&=1;\\
\text{V.2} \quad    \lambda_{\pm}(t)&=1+p_\pm t; \\  
\text{V.3} \quad    \lambda_{\pm}(t)&=1+p_{1,\pm} t+p_{2,\pm} t^2. 
\end{aligned}
\end{equation}
Matching the model with experimental data are given in Sect. \ref{sect:fresults}.

\subsection{ Full FMO model (Model B) }\label{sect:modelB}

For spin-non-flip amplitudes we use the FMO model at $t\neq 0$ \cite{MN-3}.  The only difference with  \cite{MN-3} is a notation for constants and functions. 

\begin{equation}\label{eq:bare-1}
\begin{array}{ll}
\Phi_i^{_+}(z_t,t)&=P^{(F)}_i(z_t,t)+P_i^{(1)}(z_t,t)+P_i^{(2)}(z_t,t)\\
&+P_i^{(h)}(z_t,t)+R^+_i(z_t,t),\\
%&+F^{OO}(z_t,t),\\
\Phi_i^{-}(z_t,t)&=O_i^{(M)}(z_t,t)+O_i^{(1)}(z_t,t)+O_i^{(2)}(z_t,t)\\
&+O_i^{(h)}(z_t,t)+R_i^-(z_t,t).
\end{array}
\end{equation} 
where $z_t=-1+2s/(4m^2-t)\approx 2s/(4m^2-t)$ and $i=1,5$.

In  Eq. \eqref{eq:bare-1}  $P_i^{(F)}, O_i^{(M)}$ are the Froissaron and Maximal Odderon contributions, $P_i^{(1)}, O_i^{(1)}$ are the standard (single $j$-pole) pomeron and odderon contributions and $R^+_i,  R_i^-$ are effective $f$ and $\omega$ single $j$-pole contributions, where $j$ is an angular momentum of these reggeons. $P_i^{(2)}$ and  $O_i^{(2)}$ are double $PP, OO$ and $PO$ cuts, respectively. We take into account the ''hard`` pomeron and odderon $P_i^{(h)}, O_i^{(h)}$ as well.  

We consider the model at $t\neq 0$ and at energy $\sqrt{s}> 19$ GeV, so we neglect the rescatterings  of secondary reggeons $R^+_1,  R_1^-$ with $P$ and $O$. In the considered kinematical region they are small. Besides, because $f$ and $\omega$ are effective, they  can take into account small effects from the cuts via their parameters. 

The explicit expressions for the spin-non-flip amplitudes in the  full FMO  model  are given in Appendix  \ref{sect:fmo}\,.

We used the following forms of spin-flip amplitudes for the fit:  
\begin{equation}\label{eq:Phi-5}
\begin{array}{ll}
\Phi_5^{_+}(z_t,t)&=\dfrac{\sqrt{-t}}{2m} \lambda_{+}(t)\\[3mm]
&\times \left \{ P^{(F)}_5(z_t,t)+P_5^{(eff)}(z_t,t)+R^+_5(z_t,t)\right \},\\
&\\
\Phi_5^{-}(z_t,t)&=\dfrac{\sqrt{-t}}{2m} \lambda_{-}(t)\\[3mm]
&\times \left \{O_5^{(M)}(z_t,t)+O_5^{(eff)}(z_t,t)+R_5^-(z_t,t)\right \}.\\
\end{array}
\end{equation}
For the sake of reduced number of free parameters we consider the sum of all standard pomeron (odderon) terms as one effective pomeron (odderon): 
\begin{equation}\label{eq:P,O-effective}
	\begin{aligned}
		P_1^{(eff)}(z_t,t)=P_1^{(1)}(z_t,t)+P_1^{(2)}(z_t,t)+P_1^{(h)}(z_t,t),\\
		O_1^{(eff)}(z_t,t)=O_1^{(1)}(z_t,t)+O_1^{(2)}(z_t,t)+O_1^{(h)}(z_t,t). 
	\end{aligned}
\end{equation}
Then
\begin{equation}\label{eq:Phi5-even terms}
\begin{array}{ll}
P^{(F)}_5(z_t,t)&=h_{5}e^{\beta_5{(F)}\tau _p} P^{(F)}_1(z_t,t), \\[2mm] 
 \tau _p&=2m_\pi -\sqrt{4m^2_\pi  -t},\\ [2mm]
P_5^{(eff)}(z_t,t)&=g_5^{(P)}e^{\beta_5{(P)}\tau_p}P^{(eff)}_1(z_t,t), \\[2mm] 
R_5^+(z_t,t)& =g_{5,+}e^{\beta_5^+\tau_p}R^+_1(z_t,t),\\[2mm]  
\end{array}
\end{equation} 
\begin{equation}\label{eq:Phi5-odd terms}
	\begin{array}{ll}
O_5^{(M)}(z_t,t)&=o_5e^{\beta_5{(M)}\tau_o}O^{(M)}_1(z_t,t),  \\ [2mm] 
\tau _0&=3m_\pi -\sqrt{9m^2_\pi  -t},\\[3mm]
O_5^{(eff)}(z_t,t)&=g_5^{(O)}e^{\beta_5{(O)}\tau_o}O^{(eff)}_1(z_t,t),\\[2mm]
R_5^-(z_t,t)&=g_{5,-}e^{\beta_5^-\tau_o}R^-_1(z_t,t).
\end{array}
\end{equation} 
For the functions $\lambda_{+}\pm(t)$ we have explored the same variants as used in the model A (Sect. \ref{sect:modelA}).

\renewcommand{\baselinestretch}{1.3}\normalsize
 
\begin{widetext}
	\begin{table*}%[!h]
		\caption{Data description in the Simplified FMO model (S-FMO) and original FMO model (FMO) with the three  choices of the functions $\lambda_\pm(t)$ in the spin-fip amplitudes. $N$ is the total number of experimental points in  the fit.  Comments on the number of free parameters in fits are given in Appendix \ref{sect:tabs-pars}}  	
		\begin{ruledtabular}
			\begin{tabular}{ccc|ccc|ccc}
				\hline                                                                                         
				\multirow{3}{*}{Process}   &  \multirow{3}{*} {Observable} &   \multirow{3}{*} {$N_o$}  & \multicolumn{3}{c|} {S-FMO model, $\chi^2/N_0$ } & \multicolumn{3}{c} {FMO model, $\chi^2/N_0$}\\
				\cline{4-9} 
				&&& \multicolumn{6}{c} {Variants of $\lambda_\pm(t)$ in Eqs. \eqref{eq:AP-5}, \eqref{eq:lambda} } \\  %\multicolumn{3}{c} {Variants} \\
				\cline{4-9}                                                                                                                                                
				&  &   &  V1     &       V2     &        V3                                    &   V1          &      V2     &      V3     \\
				\hline   
				$pp\to pp$          &     $\sigma_{tot}$                    &     110      &   0.899   & 0.864  & 0.865              &   0.880   &   0.890     & 0.894    \\
				$\bar pp\to \bar pp$   &     $\sigma_{tot}$            &       58       &  2.193   & 1.271   & 1.130              &   0.896     &   0.900     & 0.868   \\ 	
				$pp\to pp$          &       $\rho$                               &       67      &  1.788   & 1.586   & 1.587             &  1.563      &  1.562     & 1.564     \\ 
				$\bar pp\to \bar pp$   &       $\rho$                       &       12      &  1.267   & 0.658   &  0.595             &  0.389      &  0.379     & 0.394    \\ 
				$pp\to pp$          &      $d\sigma/dt$                      &   1701      &   1.728  & 1.630   &  1.587            &   1.522     & 1.513     & 1.514     \\ 
				$\bar pp\to \bar pp$   &      $d\sigma/dt$               &   389       &   1.359  & 0.943   &  0.919            &   1.106     & 1.063     & 1.094   \\ 
				$ pp\to pp$          &      $P(t)$                                &     49       &    0.920 & 1.742   & 1.483             &    0.908     & 0.882     & 1.105    \\ 
				\hline                                                                                                                                      
				&	$n_{par}$, number of free parameters                 &               &    35      &   37      &  38                 &    42         &   44         &   48         \\ 	
				&$\chi^2/NDF=  \chi^2/(\sum N-n_{par})$          &          &   1.648  & 1.493      & 1.449           & 1.410       & 1.405      & 1.405   \\ 	
			\end{tabular}
		\end{ruledtabular}
		\label{tab:wtab}
	\end{table*}

%\end{widetext}
%\begin{widetext}
	\begin{figure*}%[!h]
		\centering
		\includegraphics[width=1.0\linewidth]{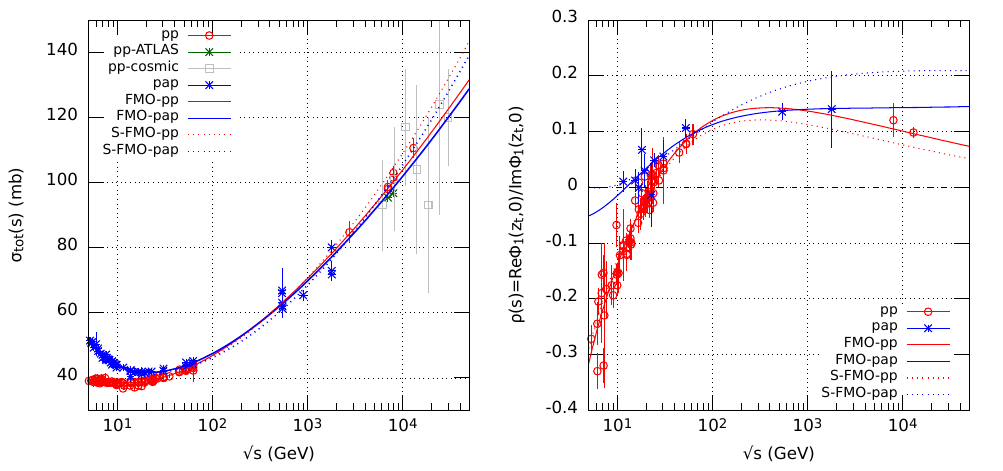}
		\caption{Total cross sections $\sigma_{tot}$ and ratio $\rho$ in the  variant V1 for models  A (dotted line) and B (solid line), $z_t(t=0)=s/2m^2-1$}
		\label{fig:sig-rho-fmo-s-fmo}
	\end{figure*}
\end{widetext}

\renewcommand{\baselinestretch}{1.0}\normalsize

\section{Results}\label{sect:fresults}

Free parameters of the models were determined from the fit to the data on $\sigma_{tot}(s), \,\, \rho(s,0), \,\, d\sigma (s,t)/dt $ and $P(t)$ in the region
\begin{equation}
	\label{eq:dataregion}
	\begin{aligned}
		&\text{for} \quad \sigma_{tot}(s), \rho(s) \, & \text{at}& \,\,5 \,\,  \text{GeV} \leq \sqrt{s}  \leq 13 \,\, \text{TeV}, \\
		&\text{for} \quad d\sigma(s,t)/dt \, & \text{at} &\,\, 19\,\, \text{GeV}  \leq \sqrt{s}  \leq 13 \,\, \text{TeV},\\ 
		&\text{for} \quad P_{pp\to pp}(t) \, & \text{at}& \,\,\sqrt{s}= 19.416,  \,\, 23.671\,\, \text{GeV}.
	\end{aligned}
\end{equation}

The data on $\sigma_{tot}$ and $\rho$ are taken from the Particle Data Group site \cite{PDG}. The recent TOTEM data \cite{TOTEM-7,TOTEM-8,TOTEM-13,TOTEM-2-76} also were added.  We did not include to the fit two ATLAS points on $\sigma_{tot}$  at 7 and 8 TeV (see discussion in \cite{MN-2}). The  data on differential cross sections were collected from the big number ot the papers from FNAL, ISR, CERN and other experimental collaborations. They can be found at the Repository for publication-related High-Energy Physics data \cite{HEPdata}. Data on polarization included in the data set for fit are taken from Refs. \cite{Fidecaro-2,Kline-1}. 

\begin{figure}[!h]
	%	\centering
	\includegraphics[width=1.0\linewidth]{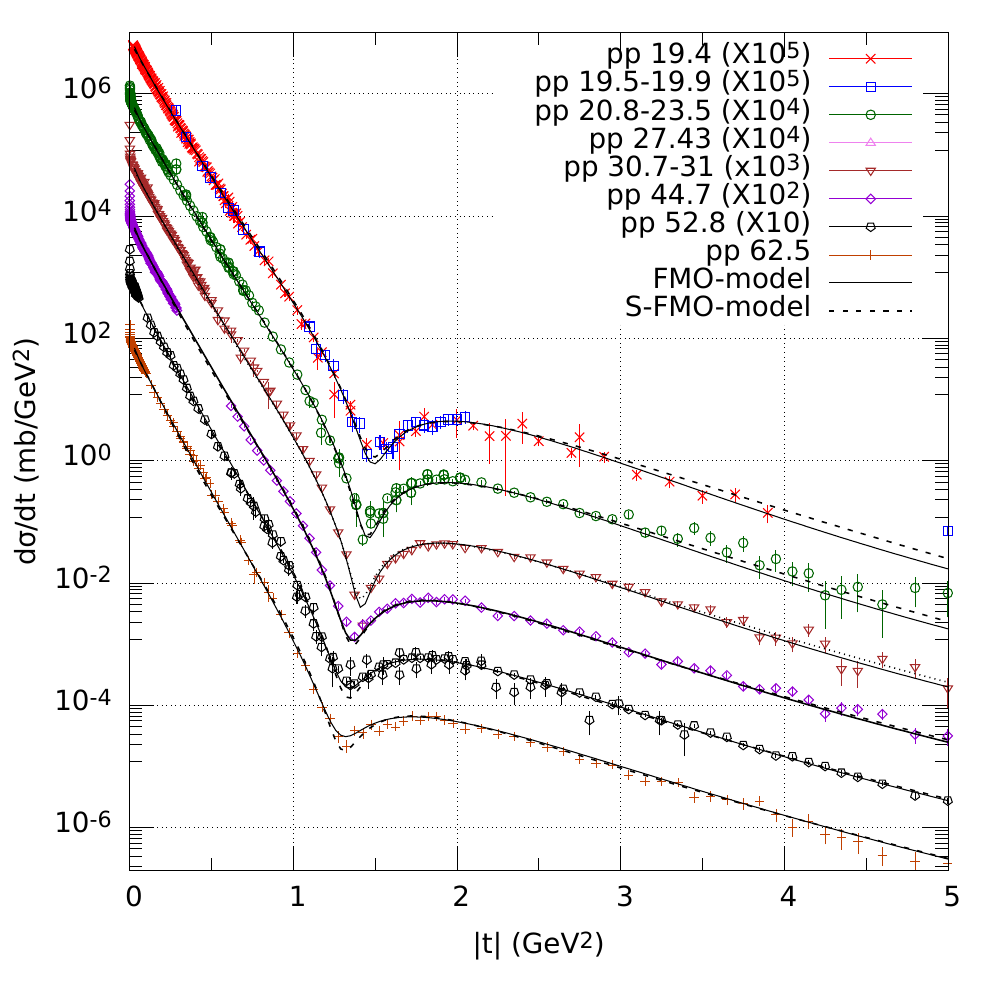}
	\caption{Differential $pp\to pp$ cross sections at the FNAL and ISR energies}
	%	 (upper panel) and at LHC energies (bottom panel)}
	\label{fig:dsdt-high-fmo-s-fmo}
\end{figure}

\begin{figure}[!h]
	%	\centering
%	\includegraphics[width=1.0\linewidth]{dsdt-high-FMO-S-FMO}
	\includegraphics[width=1.0\linewidth]{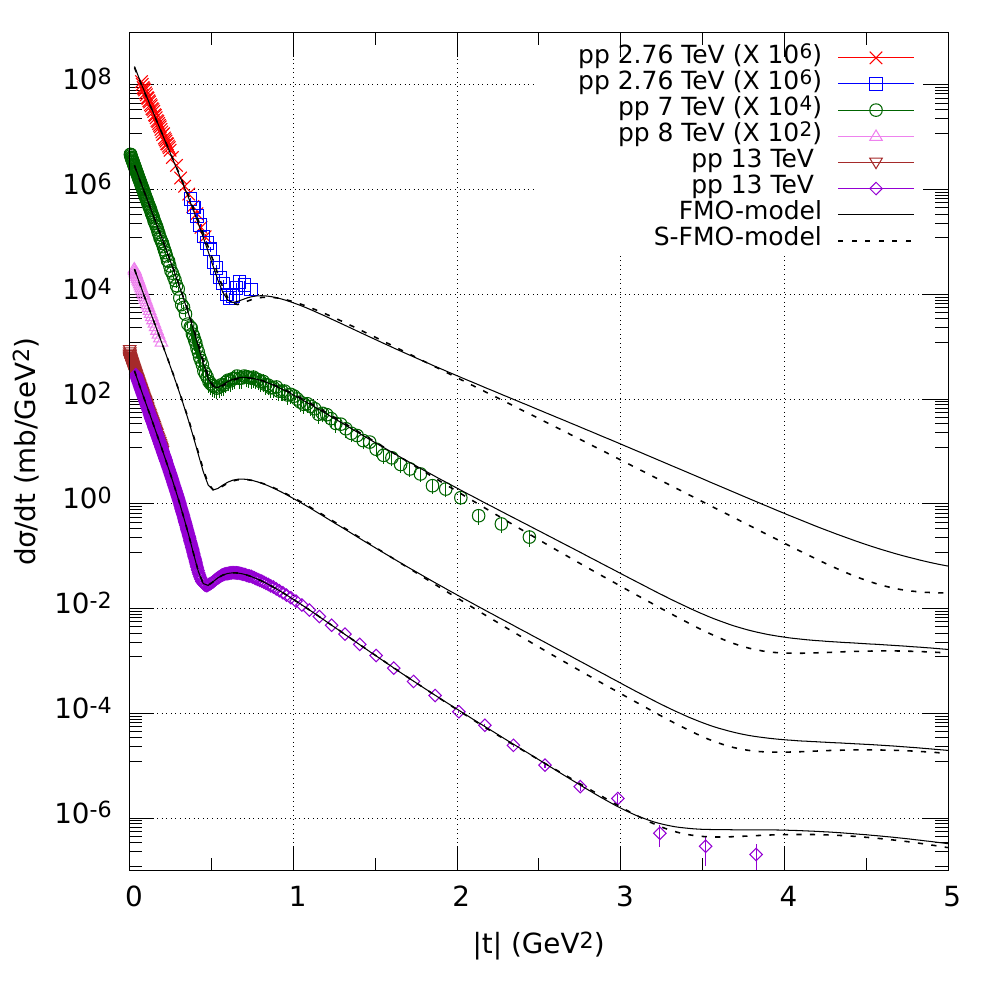}
	\caption{Differential $pp\to pp$ cross sections at the 
		%		FNAL and ISR energies (upper panel) and at 
		LHC energies} 
	%(bottom panel)}
	\label{fig:pp-TOTEM-FMO-S-FMO}
\end{figure}

\begin{figure}[!h]
	\centering
	\includegraphics[width=1.0\linewidth]{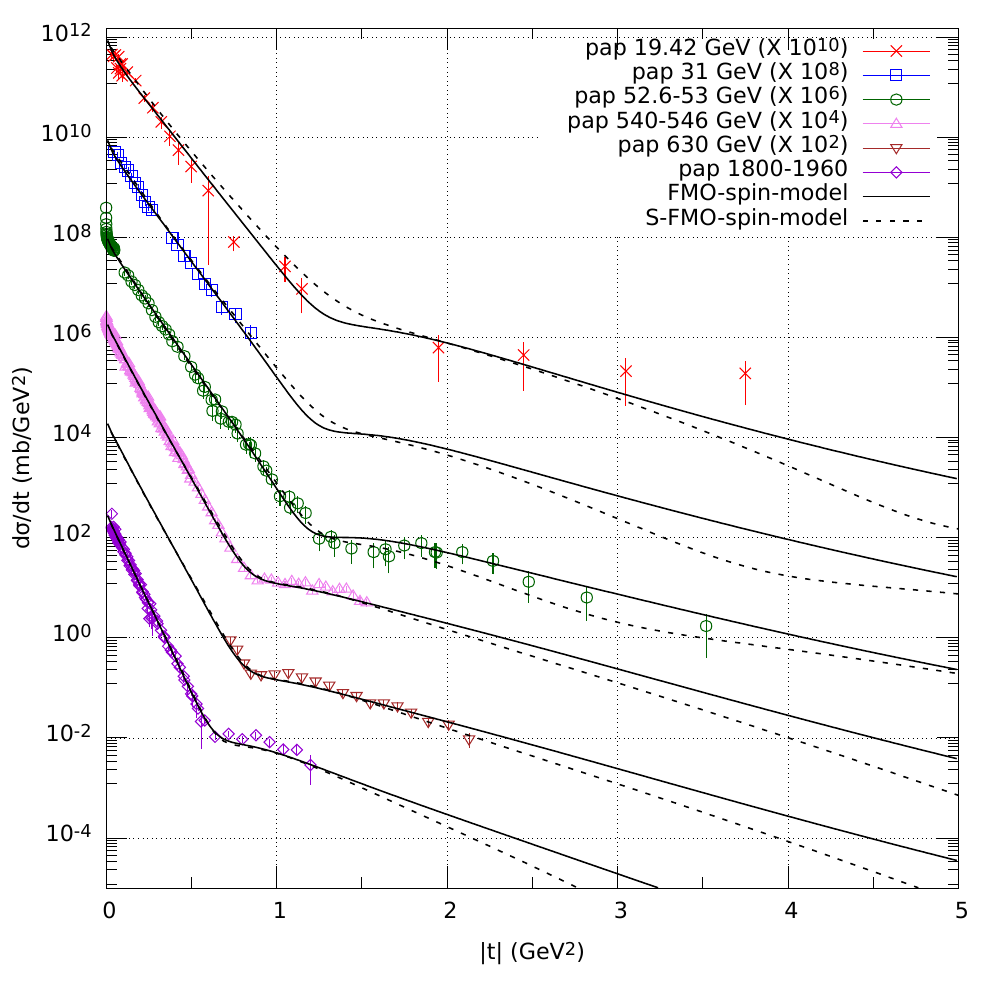}
	\caption{Differential $\bar pp\to \bar pp$ cross sections at the FNAL, ISR, SPS and Tevatron energies}
	\label{fig:pap-fmo-s-fmo}
\end{figure}

\begin{figure}[!h]
	\centering
	\includegraphics[width=1.0\linewidth]{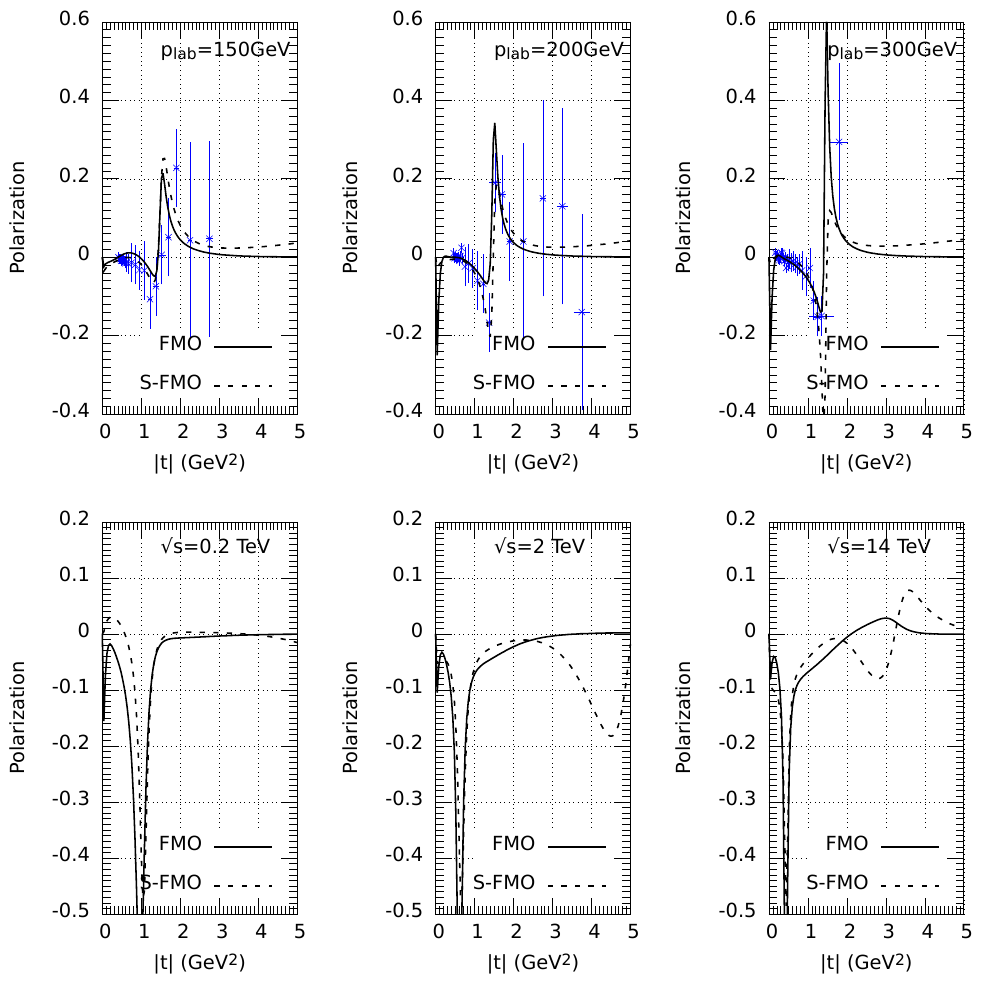}
	\caption{Polarization at the FNAL energies, and extrapolation of $P(t)$ up to $\sqrt{s}$=13 TeV. Data at $p_{lab}$=150 GeV did not included in the fit}
	\label{fig:polariz-fmo-s-fmo}
\end{figure}

Results of the fits in all considered variants of the models A and B are shown in Table \ref{tab:wtab}\,.   Because of the relatively small differences in $\chi^2$ we have plotted the Figures \ref{fig:sig-rho-fmo-s-fmo} - \ref{fig:polariz-fmo-s-fmo} 
with theoretical curves for both the models in the simplest variants V1 with $\lambda_\pm=1$. 
The values and errors of free parameters in the chosen variant  V1  are given in  Appendix \ref{sect:tabs-pars}  in Tables \ref{ltab:table-2} and Table \ref{ltab:table-3}.
One can see from the  Figures  a quite good agreement with the data on cross sections and polarization in the kinematic region \eqref{eq:dataregion}. 

We would like to note here a little bit overestimated total cross sections at the highest energies  and as consequence lower absolute values of the ratio $\rho$ in the Simplified FMO model. Additional terms, the ''hard" pomeron and odderon (exactly as it is written in the original FMO model) can fix the problem (then $\chi^2$ and curves for $\sigma_{tot}$,  $\rho$ in both the models  almost coincide). We do not discuss here this solution.

\begin{figure}[!h]
	\centering
	\includegraphics[width=1.0\linewidth]{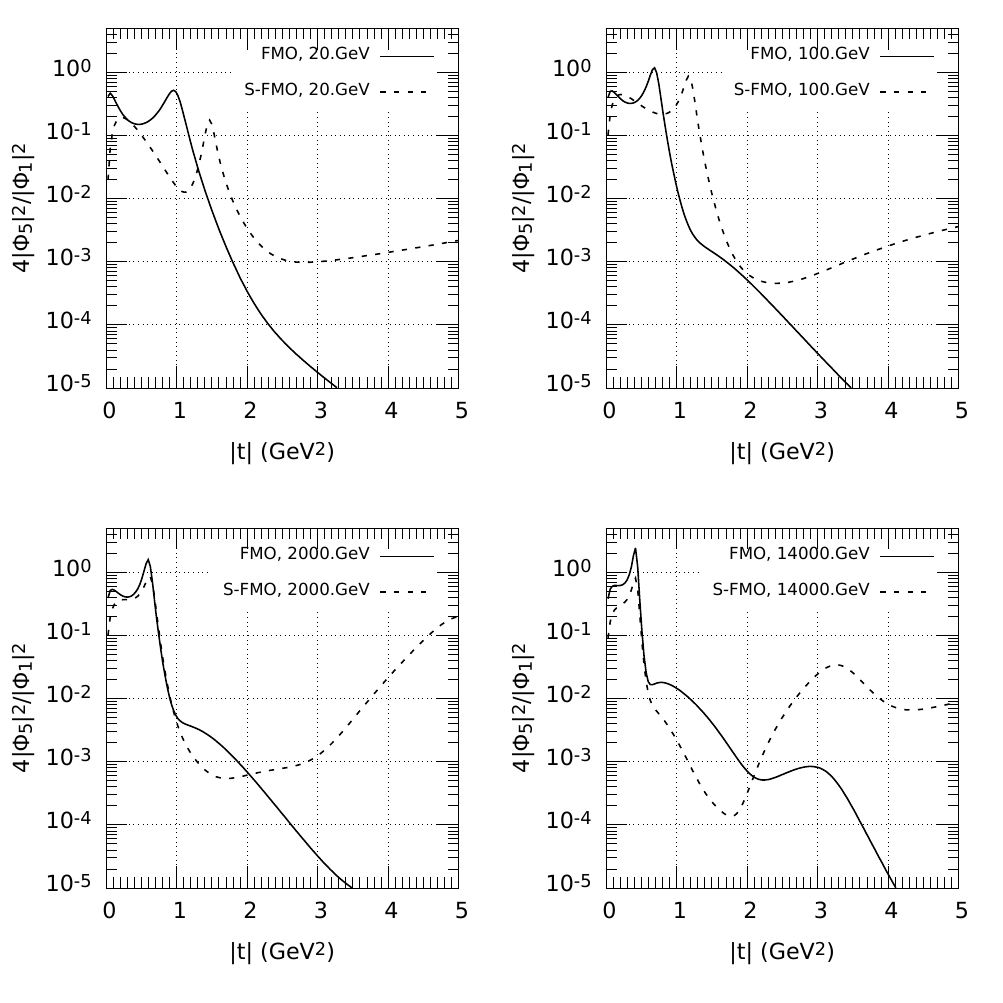}
	\caption{Ratio $4|\Phi_5|^2/|\Phi_1|^2$ of the contributions to differential $pp$ cross sections}
	\label{fig:phi5tophi1-fmo-s-fmo}
\end{figure}

\begin{figure}[!h]
	\centering
	\includegraphics[width=1.0\linewidth]{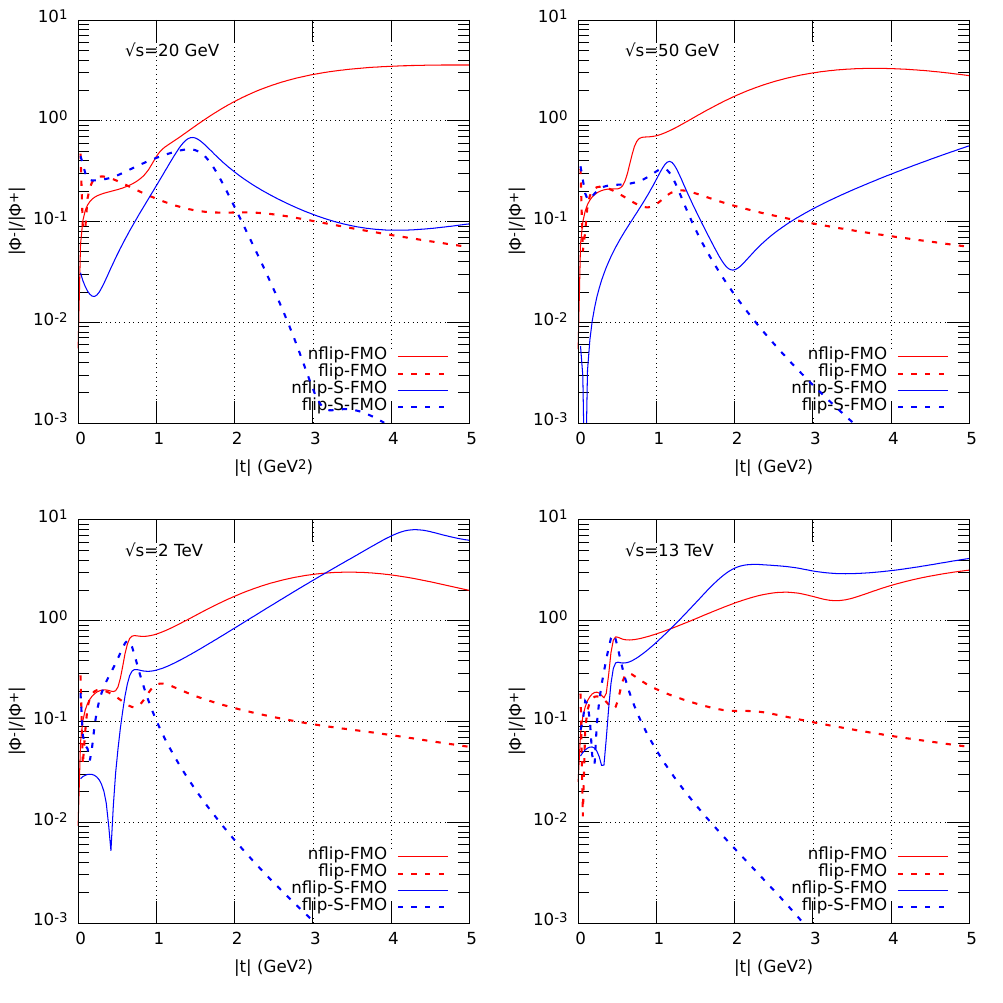}
	\caption{The ratios of crossing-odd to crossing-even parts of the spin-non-flip (solid lines) and spin-flip (dashed lines) amplitudes in the FMO and S-FMO models}
	\label{fig:ratio-min2plus-fmo}
\end{figure}

Fig. \ref{fig:phi5tophi1-fmo-s-fmo} demonstrates that the ratios  of spin-flip amplitudes to spin-non-flip ones are more important at small and middle $|t|$, decreasing with  $|t|$. The S-FMO model predicts more strong spin effects, in particular, in the region around dip in the differential cross sections and at higher $|t|$.

One can see in Fig.\;\,\ref{fig:ratio-min2plus-fmo} the Influence of odderon in the spin-non-flip and spin-flip  amplitudes in the considered models. Contribution of the Maximal  Odderon in non-flip amplitudes at not highest energies is stronger in the original FMO model (red lines) and decreases with $|t|$ slower than in the S-FMO model (blue lines). In the ''TeV`` region situation is opposite. The Maximal Odderon to Froissaron  terms ratio in the spin-flip components (dashed lines) ($|\Phi^-|_5/|\Phi^+_5|\approx 10 \text{\%}$ at the $|t|\gtrsim 1-2 \,\, \text{GeV}^2$)  in these models is higher than it was estimated in \cite{Leader} ($\sim$2\%).

The last comments in fact give the answers  to the questions addressed in Introduction about the role and magnitude of spin effects in the models within the FMO approach.

We confirm conclusions of Refs. \cite {Bultimore, Leader, DDLN, Predazzi,Selyugin-1}  obtained in more traditional pomeron and odderon models  that the crossing-even and crossing-odd spin-flip amplitudes  are important for agreement with experimental data. It would be not worse to note  that polarization in FMO model goes to 0 with rising $|t|$,  while in S-FMO  it does not vanish at least at a few GeV$^2$. 

Thus, concluding the brief discussion of the results we would like to emphasize that considered FMO models taking into account the spin-flip effects are in agreement with available experimental data.  They predict more important role of the spin-flip amplitudes at higher  energies  and momenta transferred than it is predicted by the traditional Regge pole models. Unfortunately, however,  there are no data at high energy  to confirm or deny these predictions.

{\bf Acknowledgment.} The authors thank Prof. B.~Nicolescu and Prof. S.M. Troshin for a careful reading  the manuscript, comments and useful discussions. E.M. and G.T. were  partially supported by the Project  of the National Academy of Sciences of Ukraine (0118U003197).

\appendix

\section{Simplified FMO model} \label{sect:sfmo}

The simple and dipole Pomerons (Odderons) for both components have a conventional form:
\begin{equation}\label{eq:P-S-D-1}
\begin{aligned}
P^{(S)}_{1}(s,t)&=-g^{(S)}_{1,p}\tilde s^{\alpha_p(t)}e^{\beta^{(S)}_{1,p}\tau_p}, \\  
	P^{(D)}_{1}(s,t)&=-g^{(D)}_{1,p} \xi \tilde s, ^{\alpha_p(t)}e^{\beta^{(D)}_{1,p}\tau_p},\\	 \tau_p&=2m_\pi ^2-\sqrt{4m_\pi ^2-t}, 
\end{aligned}
\end{equation}
\begin{equation}\label{eq:O-S-D-1}
\begin{aligned}
O^{(S)}_{1}(s,t)&=ig^{(S)}_{1o}\tilde s^{\alpha_o(t)}e^{\beta^{(S)}_{1,o}\tau_o},\\   
O^{(D)}_{1}(s,t&)=ig^{(D)}_{1,o}\xi 
\tilde s^{\alpha_o(t)}e^{\beta^{(D)}_{1,o}\tau_0},\\
\tau_0&=3m_\pi ^2-\sqrt{9m_\pi ^2-t}.
\end{aligned}
\end{equation}
where $\tilde s=-i(s-2m^2)/2m^2$, $\xi=\ln(\tilde s)$.

It is important that we use for single and double $j$-poles the pomeron and odderon intercepts equal to one
\begin{equation}\label{eq:P-O-alpha}
\alpha_p(t)=1+\alpha'_pt, \qquad \alpha_o(t)=1+\alpha'_ot.
	\end{equation}	
 avoiding a violation of unitarity restriction for total cross sections. 

\begin{equation}\label{eq:P-S-D-5}
\begin{aligned}
P^{(S)}_{5}(s,t)&=-g^{(S)}_{5,p}\dfrac{\sqrt{-t}}{2m}\lambda_+(t)\tilde s^{\alpha_p(t)}e^{\beta^{(S)}_{5,p}\tau_p},\\   
 P^{(D)}_{5}(s,t)&=-g^{(D)}_{5,p}\dfrac{\sqrt{-t}}{2m}\lambda_+(t)\xi \tilde s^{\alpha_p(t)}e^{\beta^{(D)}_{5,p}\tau_p},	
\end{aligned}
\end{equation}
\begin{equation}\label{eq:O-S-D-5}
\begin{aligned}
O^{(S)}_{5}(s,t)&=ig^{(S)}_{5,o}\dfrac{\sqrt{-t}}{2m}\lambda_-(t)\tilde s^{\alpha_o(t)}e^{\beta^{(S)}_{1,o}\tau_o},\\ 
O^{(D)}_{5}(s,t&)=ig^{(D)}_{5,o}\dfrac{\sqrt{-t}}{2m}\lambda_-(t)\xi  \tilde s^{\alpha_o(t)}e^{\beta^{(DS)}_{5,0}\tau_po}
\end{aligned}
\end{equation}
where $m$ is proton mass. A choice of  $\lambda_{\pm}(t)$ is discussed in the Section \ref{sect:modelA}\,.

The  tripole terms $P^{(T)}, O^{(T)}$,  containing only asymptotic main components from the Froissaron and Maximal Odderon \cite{MN-3}),  which are defined  in according with the AKM asymptotic theorem \cite{AKM}
\begin{equation}\label{eq:F-1-5}
\begin{aligned}
P^{(T)}_1(s,t)&=-g^{(T)}_{1,p}\tilde s\xi^2 \dfrac{2J_1(z_p)}{z_p}e^{\beta^{(T)}_{1,p}\tau_p},\\  P^{(T)}_5(s,t)&=-g^{(T)}_{5,p}\dfrac{\sqrt{-t}}{2m}\lambda_p(t)\tilde s\xi^2\dfrac{2J_1(z_p)}{z_p}e^{\beta^{(T)}_{5,p}\tau_p}, 
 \end{aligned}
\end{equation}
\begin{equation}\label{eq:O-1-5}
\begin{aligned}
O^{(T)}_1(s,t)&=ig^{(T)}_{1,o}\tilde s\xi^2\dfrac{2J_1(z_o)}{z_o}e^{\beta^{(T)}_{1,o}\tau_o},\\ 
O^{(T)}_5(s,t)&=ig^{(T)}_{5,o}\dfrac{\sqrt{-t}}{2m}\lambda_o(t)\tilde s\xi^2\dfrac{2J_1(z_o)}{z_o}e^{\beta^{(T)}_{1,o}\tau_o},
\end{aligned}
\end{equation}
where  $z_p=r_p\tau \xi, \quad  z_o=r_o\tau \xi, \quad \tau=\sqrt{-t/t_0}, \quad t_0=1\, \text{GeV}^2, \quad r_p, r_o$ are constants.

Contributions of the secondary reggeons, $f$ and $\omega$ have a standard form
\begin{equation}\label{eq:Rf}
\begin{aligned}
R^{(f)}_1(s,t)&=-g^{(f)}_{1}\tilde s^{\alpha_f(t)}e^{\beta^{(f)}_{1}\tau_p}, \\
R^{(f)}_5(s,t)&=-g^{(f)}_{5}\dfrac{\sqrt{-t}}{2m}\lambda_p(t)\tilde s^{\alpha_f(t)}e^{\beta^{(f)}_{5}\tau_p},\\
\alpha_f(t)&=\alpha_f(0)+\alpha'_f t,  
\end{aligned}
\end{equation}
\begin{equation}\label{eq:Romega}
\begin{aligned}
R^{(\omega)}_1(s,t)&=ig^{(\omega)}_{1}\tilde s^{\alpha_f(t)}e^{\beta^{(\omega)}_{1}\tau_o},\\ 
R^{(\omega)}_5(s,t)&=ig^{(\omega)}_{5}\dfrac{\sqrt{-t}}{2m}\lambda_o(t) \tilde s^{\alpha_\omega (t)}e^{\beta^{(\omega)}_{5}\tau_o}, \\
\alpha_\omega(t)&=\alpha_\omega (0)+\alpha'_\omega t.
\end{aligned}
\end{equation}

\section{Original FMO model}\label{sect:fmo}
We write here the explicit  spin-non-flip terms of the FMO model at $t\leq 0$ \cite{MN-3} just  for reader's convenience. The only difference of  \cite{MN-3} is a notation for constants and functions. Spin-flip terms are presented and discussed in Section \ref{sect:modelB}.

Froissaron and Maximal Odderon are written in the following form  

\begin{equation}\label{eq:MN}
\begin{array} {ll}
\dfrac{1}{iz}P^{(F)}_1(z_t,t)=h_{1,1}\xi^2\dfrac{2J_{1}(r_{+}\tau \xi)} {r_{+}\tau \xi }e^{\beta^{(F)}_{1,1}\tau_p}  \\
+h_{1,2}\xi\dfrac{\sin(r_{+}\tau \xi)}{r_+\tau \xi}e^{\beta^{(F)}_{1,2}\tau_p} +h_{1,3}J_0(r_{+}\tau \xi)e^{\beta^{(F)}_{1,3}\tau_p}  ,\\
\end{array}
\end{equation}
\begin{equation}\label{eq:MO}
\begin{array}{ll}
\dfrac{1}{z}O_1^{(M)}(z_t,t)=
o_{1,1}\xi^2\dfrac{2J_{1}(r_{-}\tau \xi)}{r_{-}\tau \xi}e^{\beta^{(O)}_{1,1}\tau_o}  \\
+ o_{1,2}\xi \dfrac{\sin(r_{-}\tau \xi)}{r_{-}\tau \xi}e^{\beta^{(OM)}_{1,2}\tau_o}
+o_{1,3}J_0(r_{-}\tau \xi)e^{\beta^{(M)}_{1,3}\tau_o} ,\\ 
\end{array}
\end{equation}
where $z=2m^2z_t, \quad \xi=\ln(-iz_t),  \quad \tau=\sqrt{-t/t_0}, \quad   t_0=1\,\,\text{GeV}^2$, $r_- =r_+ -\delta r_-,\quad \delta r_-\geq 0$. It was found in Ref. \cite{MN-3} that the best data description is achieved for minimal  allowed $\delta r_-=0$. This option in the FMO is explored in the present paper as well.  

The standard Regge pole contributions have the form
\begin{equation}\label{eq:sec Regge}
\begin{aligned}
{R_1^{(K)}}(z_t,t)&=-\binom{1}{i}
2m^2g_1^{(K)}(-iz_t)^{\alpha_K(t)}\\
&\times \left [d_ke^{b_{1,1}^{(K)}t}+(1-d_k) e^{b_{1,2}^{(K)}t}\right ], \\
\alpha_K(t)&=\alpha_K(0)+\alpha'_K t,\\ K&=P,O,f,\omega,\quad  k=p,o, \quad d_\pm =1.
\end{aligned}
\end{equation}
They take into account a possibility of a non pure exponential behaviour of the vertex functions for the standard pomeron and odderon \cite{MN-3}. 

The factor $2m^2$ is inserted in amplitudes  $R_1^{K}(z_t,t)$ in order to have the normalization for amplitudes and dimension of coupling constants (in millibarns ) coinciding with those in \cite{MN-1}. The same is made for all other amplitudes. 

We have added in the FMO  the double pomeron and odderon cuts, $PP, OO, PO$ in their usual standard form  without any new parameters as well.  Namely,
\begin{equation}\label{eq:P2}
\begin{aligned}
P_1^{(2)}(z_t,t)&=P_1^{(PP)}(z_t,t)+P_1^{(OO)}(z_t,t),\\
O_1^{(2)}(z_t,t)&=P_1^{(PO)}(z_t,t),
\end{aligned}
\end{equation}

\renewcommand{\baselinestretch}{1.3}\normalsize

\begin{equation}\label{eq:PP}
\begin{array}{ll}
P_1^{(PP)}(z_t,t)&=-i\dfrac{2m^2(z_tg_1^{(P)})^2}{16\pi s\sqrt{1-4m^2/s}} \left \{ \dfrac{d_p^2}{2B_1^p}\exp(tB_1^p/2)\right .\\[3mm]
&
+\dfrac{2d_p(1-d_p)} {B_1^p+B_2^p}\exp\left (t\dfrac{B_1^pB_2^p}{B_1^p+B_2^p} \right ) \\[3mm]
&\left .+\dfrac{(1-d_p)^2} {2B_2^p}\exp(tB_2^p/2) \right \} ,
\end{array}
	\end{equation}
\begin{equation}\label{eq:OO}
\begin{array}{ll}
P_1^{OO}(z_t,t)&=-i\dfrac{2m^2(z_tg_1^{(O)})^2}{16\pi s\sqrt{1-4m^2/s}} 
\left \{ \dfrac{d_o^2}{2B_1^o}\exp(tB_1^o/2)\right . \\[3mm]
&+\dfrac{2d_o(1-d_o)} {B_1^o+B_2^o}\exp\left (t\dfrac{B_1^oB_2^o}{B_1^o+B_2^o} \right )\\[3mm]
&+\left .\dfrac{(1-d_o)^2} {2B_2^o}\exp(tB_2^o/2) \right \} ,
\end{array}
\end{equation}
\begin{equation}\label{eq:PO}
\begin{array}{ll}
P_1^{PO}(z_t,t)&=\dfrac{2m^2z_t^2g_1^{(P)} g_1^{(O)}} {16\pi s\sqrt{1-4m^2/s}}\\[3mm]
&\times \left \{
\dfrac{d_pd_o}{B_1^p+B_1^o}\exp\left (t\dfrac{B_1^pB_1^o}{B_1^p+B_1^o}\right )\right .\\[3mm]
&+ \dfrac{d_p(1-d_o)}{B_1^p+B_2^o}\exp\left (t\dfrac{B_1^pB_2^o}{B_1^p+B_2^o}\right)\\[3mm]
& +\dfrac{(1-d_p)d_o}{B_2^p+B_1^o}\exp\left(t\dfrac{B_2^pB_1^o}{B_2^p+B_1^o}\right )\\[3mm]
&+\left . \dfrac{(1-d_p)(1-d_o)}{B_2^p+B_2^o}\exp\left(t\dfrac{B_2^pB_2^o}{B_2^p+B_2^o}\right) \right \}  
\end{array}
\end{equation}
where $B_k^{p,o}=b_{1,k}^{P.O}+\alpha'_{P,0}\ln(-iz_t),\quad k=1,2 , \quad b_{1,k}^{P,O}$ are the constants from single pomeron and odderon contributions.

In \cite{MN-3} it was noted  that for a better description of the data it is advisable  to add to the amplitudes the  contributions that mimic some  properties of ''hard`` pomeron ($P^h$) and  odderon ($O^h$). We take them in the simplest form
\begin{equation}
\begin{aligned}
P_1^{(h)}(t)&=i2m^2z_t\dfrac{g_{h,p}}{(1-t/t_{p,h})^ 4},\\
%\begin{equation}
O_1^{(h)}(t)&=2m^2z_t\dfrac{g_{h,o}}{(1-t/t_{o,h})^{4}}.
\end{aligned}
\end{equation}

\section{Parameters in the models}\label{sect:tabs-pars} 

The number of free parameters is varied in various fits, because we put a reasonable (in our opinion) limits for  slopes in the exponential vertexes ($0\leq b, \beta \leq 20\,\text{GeV}^{-1}$)  and for those of  trajectories  ($0.8\leq \alpha'_{f, \omega} \leq 1.1\,\text{GeV}^{-2}$). If during fitting such a parameter goes to the limit  value, then it is fixed at the corresponding limit. 

\renewcommand{\baselinestretch}{1.3}\normalsize

\begin{longtable}{cccc dp{1.cm} dp{1.cm}}[H] 
\endfirsthead
\endhead
\caption{Parameters of the \textbf{Simplified FMO model (A)}} \\
	\hline
	Name       & Dimension   &  Value  	&      Error \\
	\hline
	\endfirsthead
	\multicolumn{4}{r}
	{\tablename\ \thetable\ -- \textit{Continued from previous page}} \\
	\hline
	Name    &    Dimension   & Value  	&      Error \\
	\hline
	\endhead
	\hline \multicolumn{4}{r}{\textit{Continued on next page}} \\
	\endfoot
	\hline
	\endlastfoot
	\hline
$g^{(T)}_{1,p}$ &  mb                    &  0.29343E+00 & 0.22854E-03\\
$\beta^{(T)}_{1,p}$ &  GeV$^{-1}$        &  0.51823E+01 &  0.25651E-02\\
$r_{p}$&                                 &  0.29661E+00 &  0.57101E-04\\
$\alpha'_{p}$ &  GeV$^{-2}$              &  0.16361E+00 &  0.39431E-03\\
$g^{(D)}_{1,p}$ &  mb                    & -0.76106E+00 &  0.11876E-02\\
$\beta^{(D)}_{1,p}$ &  GeV$^{-1}$        &  0.19434E+01 &  0.32444E-02\\
$g^{(S)}_{1,p}$ &  mb                    &  0.27402E+02 &  0.40164E-01\\
$\beta^{(S)}_{1,p}$ &  GeV$^{-1}$        &  0.41323E+01 &  0.38042E-02\\
$g^{(T)}_{1,o}$ &  mb                    & -0.37472E-01 &  0.22257E-03\\
$\beta^{(T)}_{1,o}$ &  GeV$^{-1}$        &  0.33673E+01 &  0.93166E-02\\
$r_{o}$&                                 &  0.27253E+00 &  0.24894E-03\\
$\alpha'_{o}$ &  GeV$^{-2}$              &  0.12120E-01 &  0.45999E-03\\
$g^{(D)}_{1,o}$ &  mb                    &  0.20536E+00 &  0.25831E-02\\
$\beta^{(D)}_{1,o}$ & GeV$^{-1}$         &  0.65245E+01 &  0.35911E-01\\
$g^{(S)}_{1,o}$ &  mb                    &  0.13783E+01 &  0.14165E-01\\
$\beta^{(S)}_{1,o}$ &  GeV$^{-1}$        &  0.34116E+01 &  0.12406E-01\\
$\alpha_{f}(0)$&                         &  0.76576E+00 &  0.15048E-02\\
$\alpha'_{f}$ &  GeV$^{-2}$              &  0.80000E+00 &  fixed at limit\\
$g^{{(f)}}_{1}$ &  mb                    &  0.35167E+02 &  0.24485E+00\\
$\beta^{(f)}_{1}$ &  GeV$^{-1}$          &  0.00000E+00 & fixed at limit\\
$\alpha_{\omega}(0)$&                    &  0.61636E+00 &  0.58521E-02\\
$\alpha'_{\omega}$ &  GeV$^{-2}$         &  0.80000E+00 &  fixed at limit\\
$g^{{(\omega)}}_{1}$ &  mb               &  0.24366E+02 &  0.35434E+00\\
$\beta^{(\omega)}_{1}$ &  GeV$^{-1}$     &  0.20000E+02 &  fixed at limi\\
$g^{(T)}_{5,p}$ &  mb                    &  0.98237E+00 &  0.29269E-02\\
$\beta^{(T)}_{5,p}$ &  GeV$^{-1}$        &  0.97449E+01 &  0.11045E-01\\
$g^{(D)}_{5,p}$ &  mb                    & -0.23787E+02 &  0.71732E-01\\
$\beta^{(D)}_{5,p}$ &  GeV$^{-1}$        &  0.12183E+02 &  0.11698E-01\\
$g^{(S)}_{5,p}$ &  mb                    &  0.29036E+03 &  0.70315E+00\\
$\beta^{(S)}_{5,p}$ &  GeV$^{-1}$        &  0.97486E+01 &  0.80005E-02\\
$g^{(T)}_{5,o}$ &  mb                    &  0.57721E-03 &  0.57266E-03\\
$\beta^{(T)}_{5,o}$ &  GeV$^{-1}$        &  0.26306E+01 &  0.96362E+00\\
$g^{(D)}_{5,o}$ &  mb                    &  0.77319E+01 &  0.13988E+00\\
$\beta^{(D)}_{5,o}$ &  GeV$^{-1}$        &  0.14185E+02 &  0.63904E-01\\
$g^{(S)}_{5,o}$ &  mb                    & -0.69382E+02 &  0.14333E+01\\
$\beta^{(S)}_{5,o}$ &  GeV$^{-1}$        &  0.12588E+02 &  0.79085E-01\\
$g^{{(f)}}_{5}$ &  mb                    & -0.58675E+03 &  0.12276E+02\\
$\beta^{(f)}_{5}$ &  GeV$^{-1}$          &  0.92251E+01 &  0.12837E+00\\
$g^{{(\omega)}}_{5}$ &  mb               &  0.29637E+02 &  0.17742E+01\\
$\beta^{(\omega)}_{5}$ &  GeV$^{-1}$     &  0.00000E+00 &  fixed at limit\\
%	\hline
\label{ltab:table-2}
\end{longtable}

\begin{longtable}{cccc dp{1.cm} dp{1.cm}}[H] 
	\endfirsthead
	\endhead
	\caption{Parameters of the \textbf{FMO model (B)}}\\
	\hline
	Name       & Dimension   &     Value  	&      Error \\
	\hline
	\endfirsthead
	\multicolumn{4}{r}
	{\tablename\ \thetable\ -- \textit{Continued from previous page}} \\
	\hline
	Name    &    Dimension   & Value  	&      Error \\
	\hline
	\endhead
	\hline \multicolumn{4}{r}{\textit{Continued on next page}} \\
	\endfoot
	\hline
	\endlastfoot
	\hline
$ h_{1,1}$ & mb                              &  0.14330E+00   &  0.11569E-03\\
$ h_{1,2}$ & mb                              &  0.29852E+01   &  0.54259E-02\\
$ h_{1,3}$ & mb                              &  0.90696E+01   &  0.10925E+00\\
$ r_{+}$   &                                      &  0.26288E+00   &  0.54777E-04\\
$\beta^{(F)}_{1,1}$ & GeV$^{-1}$             &  0.23291E+01   &  0.13574E-02\\
$\beta^{(F)}_{1,2}$ & GeV$^{-1}$             &  0.39343E+01   &  0.44429E-02\\
$\beta^{(F)}_{1,3}$ & GeV$^{-1}$             &  0.15649E+02   &  0.67054E+00\\
$o_{1,1}$   &  mb                           &  -0.42452E-01   &  0.16579E-03\\
$o_{1,2}$   &  mb                           &  0.94686E+00   &  0.17991E-01\\
$o_{1,3}$   &  mb                           & -0.14200E+02   &  0.10239E+00\\
$ r_{-}$    &                                    &  0.26288E+00   &  fixed:  $r_-=r_+$ \\
$\beta^{(O)}_{1,1}$ & GeV$^{-1}$             &  0.15717E+01   &  0.34223E-02\\
$\beta^{(O)}_{1,2}$ & GeV$^{-1}$             &  0.56167E+01   &  0.14887E+00\\
$\beta^{(O)}_{1,3}$ & GeV$^{-1}$             &  0.31265E+01   &  0.93369E-02\\
$\alpha'_{p}$ & GeV$^{-2}$                   &  0.17820E+00   &  0.15904E-03\\
$g^{(P)}_{1}$ & mb                           &  0.71055E+02   &  0.56702E-01\\
$d_{p}$ &       mb                           &  0.73915E+00   &  0.74388E-03\\
$b^{(P)}_{1,1}$ & GeV$^{-2}$                 &  0.52382E+01   &  0.49458E-02\\
$b^{(P)}_{1,2}$ & GeV$^{-2}$                 &  0.21387E+01   &  0.29922E-02\\
$\alpha'_{o}$ &   GeV$^{-2}$                 &  0.41252E-02   &  0.15522E-03\\
$g^{(Od)}_{1}$ & mb                          &  0.27241E+02   &  0.36231E-01\\
$d_{o}$ &            mb                      &  0.75363E+00   &  0.70380E-03\\ 
$b^{(Od)}_{1,1}$ & GeV$^{-2}$                &  0.62168E+01   &  0.88381E-02\\
$b^{(Od)}_{1,2}$ & GeV$^{-2}$                &  0.21600E+01   &  0.26952E-02\\
$\alpha_{f}(0)$ &                            &  0.76841E+00   &  0.10061E-02\\
$\alpha'_{f}$ & GeV$^{-2}$                   &  0.11000E+01   &  fixed at limit\\
$ g^{(f)}_{1}$ & mb                          &  0.70300E+02   &  0.24332E+00\\
$b_{1,f}$ & GeV$^{-2}$                        &  0.11101E+02   &  0.28189E+00\\
$\alpha_{\omega}(0)$ &                       &  0.31275E+00   &  0.11785E-01\\
$\alpha'_{\omega}$ & GeV$^{-1}$              &  0.11000E+01   &  fixed at limit\\
$g^{(\omega)}_{1}$ & mb                      &  0.34042E+02   &  0.11844E+01\\
$b_{1,\omega}$ & GeV$^{-2}$                  &  0.14447E+01   &  0.22569E+00\\
$ g_{p,h}$ & mb                              & -0.73998E+02   &  0.45714E-01\\
$t_{p,h}$ & GeV$^{2}$                        &  0.38841E+00   &  0.14919E-03\\
$g_{o,h}$ & mb                               &  0.32403E+02   &  0.42717E-01\\
$t_{o,h}$ & GeV$^{2}$                        &  0.61854E+00   &  0.51543E-03\\
$ h_{5}$ & mb                                &  0.38020E+01   &  0.53841E-02\\
$ \beta^{(F)}_{5}$ & GeV$^{-1}$              &  0.31399E+01   &  0.49814E-02\\
$o_{5}$ & mb                                 &  0.38312E+01   &  0.15409E+00\\
$\beta^{(O)}_{5}$ & GeV$^{-1}$               &  0.66434E+01   &  0.13773E+00\\
$g_{5,p}$ & mb                               &  0.28732E+01   &  0.34589E+00\\
$\beta_{5,p}$ & GeV$^{-1}$                   &  0.13172E+02   &  0.69818E+00\\
$ g_{5,o}$ & mb                              &  0.18954E+02   &  0.16931E+01\\
$\beta_{5,o}$ & GeV$^{-1}$                   &  0.20000E+02   &  fixed at limit\\
$g_{5,f}$ & mb                               &  0.29711E+01   &  0.17843E+00\\ 
\samepage
$\beta_{5,f}$ & GeV$^{-1}$                   &  0.00000E+00   &  fixed at limit\\  \samepage
$ g_{5,\omega}$ & mb                         & -0.15919E+02   &  0.13466E+01\\ 
$ \beta_{5,\omega}$ & GeV$^{-1}$ &             0.00000E+00   &  fixed at limit
%\hline
\label{ltab:table-3}
\end{longtable}
	\renewcommand{\baselinestretch}{1.0}	

\end{document}